\documentclass[pra,aps,reprint,showpacs,amsmath]{revtex4-1}

\usepackage{bm}
\usepackage{graphicx}
\usepackage{dsfont}

\newcommand{\bra}[1]{\langle #1 | \,}
\newcommand{\ket}[1]{\, | #1 \rangle}
\newcommand{\braket}[2]{\langle #1 | #2 \rangle}
\newcommand{\expv}[1]{\langle #1 \rangle}

\newcommand{\Om}{\Omega}
\newcommand{\ga}{\gamma}
\newcommand{\Ga}{\Gamma}

\newcommand{\De}{\Delta}

\newcommand{\bx}{\mathbf{x}}
\newcommand{\db}{d_{\mathrm{b}}}
\newcommand{\nR}{n_{\mathrm{R}}}
\newcommand{\pR}{p_{\mathrm{R}}}
\newcommand{\hlf}{\frac{1}{2}}
\newcommand{\mc}[1]{\mathcal{#1}}
\newcommand{\sig}{\hat{\sigma}}
\newcommand{\Sig}{\hat{\Sigma}}
\newcommand{\hrho}{\hat{\rho}}
\newcommand{\hL}{\hat{L}}

\begin{document}

\title{Dynamics and equilibration of Rydberg excitations 
in dissipative atomic ensembles}

\author{David Petrosyan}
\affiliation{Institute of Electronic Structure and Laser, 
FORTH, GR-71110 Heraklion, Crete, Greece}
\date{\today}

\begin{abstract}
We study resonant optical excitations of strongly-interacting 
Rydberg states of atoms in the presence of relaxations. 
We employ the quantum stochastic (Monte Carlo) wavefunctions
to simulate the dissipative dynamics of tens of atoms in  
two-dimensional lattices. We show that under typical 
experimental conditions involving slow Rydberg state decay and
sizable relaxation of atomic coherences, on the time scale of 
several $\mu$s the atomic ensemble approaches a stationary state 
in which much of the quantum correlations between the atoms 
have decayed away. The steady state, however, exhibits strong 
classical correlations of Rydberg excitation probabilities.
\end{abstract}


\maketitle


Atoms in high-lying Rydberg states interact with each other via a 
long-range van der Waals (vdW) potential \cite{RydAtoms,rydrevRMP,rydrevJOSAB}
which is many orders of magnitude stronger than the interaction 
potential between ground-state atoms at $\mu$m distances. 
In dense atomic ensembles ($\rho_{\mathrm{at}} \gtrsim 10^{12}\:$cm$^{-3}$), 
resonant optical excitations of multiple Rydberg atoms are then 
strongly suppressed \cite{Tong2004,Singer2004,Vogt2006,Heidemann2007,Low2009} 
due to the interaction induced level shifts. In a small volume, 
where the interatomic interaction energies exceed the excitation 
linewidth of the Rydberg state, a single Rydberg atom blocks 
the excitation of all the other atoms \cite{Lukin2001,Robicheaux2005,%
Stanojevic2009,UrbanGaetan2009,Dudin2012NatPh,DPKM2013}. 
Larger atomic ensembles can accommodate more Rydberg excitations 
whose number exhibits reduced fluctuations \cite{Raithel2005,Viteau2012}. 
Recent experiments have spectacularly demonstrated Rabi oscillations
of single Rydberg excitations in blockaded ensembles of atoms 
\cite{Dudin2012NatPh}, and spatial ordering of multiple Rydberg 
excitations in larger ensembles of atoms in two-dimensional (2D)
optical lattices \cite{Schauss2012}. 

In most experiments
\cite{Singer2004,Heidemann2007,Low2009,UrbanGaetan2009,Dudin2012NatPh,%
Raithel2005,Viteau2012,Schauss2012,Schwarzkopf2011,Johnson2008,Raitzsch2008}, 
the Rydberg state is resonantly coupled to the atomic ground state 
by a two-photon transition via a non-resonant intermediate state. 
With the pulsed laser excitation, coherent dynamics of the many-atom system 
is observed on the $\mu$s time scale \cite{Dudin2012NatPh,Schauss2012}. 
For comparison, typical lifetime of the highly excited Rydberg state 
with the principal quantum number $n \gtrsim 40$ is $\tau_r \sim 20\:\mu$s,
while the intermediate state decay, two-photon laser linewidth, 
stray electromagnetic fields, atomic motion and collisions
dephase the atomic polarization resulting in the coherence time 
of $\tau_{z} \lesssim 10\:\mu$s. It is therefore important to 
quantify the role of relaxations in the dynamical evolution
of the system and its equilibration, and to characterize the 
properties of the resulting stationary state which may indeed
be different from the transient or time-averaged state of a fully
coherent (unitary) system. 
    
Exact treatment of strongly-interacting many-body systems is 
computationally demanding---the dimension of Hilbert space for 
$N$ two-level atoms is $2^N$---and even more so for dissipative 
systems. It requires the solution of $2^{2N}$ density matrix 
equations which in practice is restricted to $N \lesssim 12$ 
atoms due to computer memory limitations. An alternative but
equivalent approach employs the quantum stochastic (Monte Carlo) 
wavefunctions, which amounts to propagating $2^N$ amplitude
equations, interrupted by quantum jumps, and then averaging over
many independent trajectories. The quantum Monte Carlo simulations
can thus deal with about double the number of atoms $N$ but require
longer simulation times (or parallel computation) to generate
many trajectories for good averaging. Furthermore, limiting
the maximal number of excitations, which is justified by the 
Rydberg blockade, can greatly reduce the required Hilbert space
dimension permitting an approximate treatment of many more atoms
\cite{Viteau2012,Schauss2012,Garttner2012}.

This paper presents the results of quantum Mote Carlo simulations
for Rydberg excitation of several tens of atoms in 2D lattices. 
Typical experimental values 
\cite{Schauss2012} for the Rabi frequency of the driving field and 
the relaxation rates of atomic population and coherence are used in 
the simulations and the resulting dynamics is compared to that 
of a unitary (dissipationless) system. When all the atoms are within 
the blockade volume, the system undergoes damped Rabi oscillations 
between the collective ground and single Rydberg excitation 
states, and the final stationary state of the system contains
a single Rydberg excitation with close to unit probability \cite{DPMHMF2013}.
But when the system size is larger than the blockade volume,
the weaker interactions between distant atoms cannot block
multiple Rydberg excitations; instead they dephase the Rabi oscillations
even in the unitary system. Including relaxations with rates
comparable to the single-atom Rabi frequency steers the
system towards a stationary state in which the quantum 
correlations between the atoms have largely decayed away. 
Strong classical correlations, however, persist, as the steady 
state is characterized by very small fluctuations of the
number of excitations which is consistent with the tight
spatial ordering of Rydberg excitations \cite{DPMHMF2013,Schauss2012}.


Let us now turn to the quantitative description of the system.
A 2D lattice of $N$ atoms is irradiated by a uniform laser field 
which resonantly drives the transition from the atomic ground state $\ket{g}$
to the highly excited Rydberg state $\ket{r}$ with Rabi frequency $\Om$.
The atom-field interaction Hamiltonian reads
\[
\mc{V}_{\mathrm{af}}^j =  - \hbar \Om (\sig_{rg}^j + \sig_{gr}^j) ,
\]
where $\sig_{\mu \nu}^j \equiv \ket{\mu}_{jj}\bra{\nu}$ 
is the transition operator for atom $j$. 
The relaxation processes affecting each atom include 
the population decay of the excited state $\ket{r}$ with 
rate $\Ga_r$, and the decay of atomic coherence $\sig_{rg}$
with rate $\Ga_z$. These processes are described by  
Liouvillians in the Lindblad form 
\begin{equation}
\mc{L}^j \hrho = \hlf [2 \hL^j \hrho \hL^{j\dagger} 
- \{ \hL^{j\dagger} \hL^j , \hrho \}] ,
\end{equation}
where $\hrho$ is the density operator of the system, while 
the generators for the population and coherence decay are
given, respectively, by $\hL_r^j = \sqrt{\Ga_r} \sig_{gr}^j$ and
$\hL_z^j = \sqrt{\Ga_z} (\sig_{rr}^j -\sig_{gg}^j)$. For an isolated 
atom, the (steady-state) excitation linewidth of state $\ket{r}$
is $w \simeq 2 \Om \sqrt{\ga_{rg}/\Ga_r}$ with 
$\ga_{rg} \equiv \hlf \Ga_{r} + 2 \Ga_{z}$ and assuming
$\Om^2 > \Ga_r \ga_{rg}$ \cite{DPMHMF2013,PLDP2007}.

We next include the interatomic interactions.
The vdW potential between a pair of atoms $i$ and $j$ at 
positions $\bx_i$ and $\bx_j$ induces the level shift 
$\Delta(\bx_i -\bx_j) = C_{6} |\bx_i -\bx_j|^{-6}$
of state $\ket{r_i r_j}$, where $C_6$ is the vdW  coefficient \cite{rydcalc}.
The corresponding atom-atom interaction Hamiltonian reads
\[
\mc{V}_{\mathrm{aa}}^{ij} = \hbar  \sig_{rr}^i \Delta(\bx_i -\bx_j) \sig_{rr}^j .
\]
When the vdW level shift is larger than the Rydberg state excitation
linewidth, $\De \gtrsim w$, an atom in state $\ket{r}$ blocks 
the excitation of another atom \cite{Lukin2001,rydrevJOSAB,rydrevRMP}.
We can then define the blockade distance $\db$ via $\De(\db) = w$, 
which yields $\db \equiv \sqrt[6]{C_p/w}$.

The total Hamiltonian for $N$ atoms is given by 
\begin{equation}
\mc{H} = \sum_{j}^N \mc{V}_{\mathrm{af}}^j 
+ \sum_{i<j}^N \mc{V}_{\mathrm{aa}}^{ij} .
\end{equation}
To simulate the dissipative dynamics of the many-body system, 
we employ the quantum Monte Carlo wavefunctions \cite{qjumps,PLDP2007}.
In such simulation, the state of the system $\ket{\Psi}$ evolves 
according to the Schr\"odinger equation 
$\partial_t \ket{\Psi} = -\frac{i}{\hbar} \tilde{\mc{H}} \ket{\Psi}$ with 
an effective Hamiltonian 
\begin{equation}
\tilde{\mc{H}} = \mc{H} - \frac{i}{2} \hbar \hL^2 ,
\end{equation}
where 
\[
\hL^2 \equiv \sum_j (\hL_r^{j\dagger} \hL_r^j + \hL_z^{j\dagger} \hL_z^j )
= \sum_j \Ga_r \sig^j_{rr} + \Ga_z \mathds{1}
\]
is the non-Hermitian part which does not preserve the norm of the wavefunction 
$\ket{\Psi}$ during the evolution. The evolution is interrupted by random 
quantum jumps $\ket{\Psi} \to \hL_{r,z}^j \ket{\Psi}/P_{r,z}^j$ 
with probabilities determined by the corresponding weights 
$P_{r,z}^j \equiv \bra{\Psi} \hL_{r,z}^{j\dagger} \hL_{r,z}^j \ket{\Psi} $.
In a single quantum trajectory, the normalized wavefunction 
of the system at any time $t$ is given by
$\ket{\bar{\Psi}(t)} = \ket{\Psi(t)}/\sqrt{\braket{\Psi(t)}{\Psi(t)}}$.
The approximate density operator of the system is then obtained by
averaging over many $M$ independently simulated trajectories,
$\hrho(t) = \frac{1}{M} \sum_m^M \ket{\bar{\Psi}_m(t)} \bra{\bar{\Psi}_m(t)}$.
On the other hand, the steady-state density operator can be approximated as 
a long-time average over a single trajectory,
$\hrho(\infty) = \lim_{t \to \infty} \frac{1}{t-t_0}
\int_{t_0}^t d t \ket{\bar{\Psi}(t)} \bra{\bar{\Psi}(t)}$.
 
The mean number of Rydberg excitations within an ensemble of $N$ atoms
is $\expv{\nR} = \expv{\sum_j^N \sig_{rr}^j}$, while the probabilities 
$\pR(n) = \expv{\Sig_r^{(n)}}$ of $n$ excitations are defined through 
the corresponding projectors $\Sig_r^{(0)} \equiv \prod_{j}^N \sig_{gg}^j$,
$\Sig_r^{(1)} \equiv \sum_{j}^N \sig_{rr}^j \prod_{i \neq j}^N \sig_{gg}^j$,
etc. Obviously $\expv{\nR} = \sum_n n \, \pR(n)$, while 
$\sig_{gg}^i + \sig_{rr}^i = \mathds{1} \; \forall \; i \in [1,N]$. 
To quantify the probability distribution of Rydberg excitations,
we use the Mandel $Q$ parameter \cite{MandelQ}
\[
Q \equiv \frac{\expv{\nR^2} - \expv{\nR}^2}{\expv{\nR}} - 1 ,
\]
where $\expv{\nR^2} = \sum_n n^2 \, \pR(n)$. A Poissonian 
distribution $\pR(n) =  \expv{\nR}^n e^{-\expv{\nR}}/n!$
leads to $Q=0$, while $Q < 0$ corresponds to sub-Poissonian
distribution, with $Q=-1$ attained for a definite number $n$ of
excitations, $\pR(n) = 1$.   


We have performed numerical simulations for various number of atoms $N$ 
arranged in 2D circular volumes (disks) of different diameter $d \leq 2\db$.
In the dynamical simulations, we typically generate $M \sim 200$ 
trajectories for smooth averaging. 
We truncate the Hilbert space by limiting the maximal number of Rydberg 
excitations, but verify the convergence by including more excitations, 
at the expense of less accurate averaging 
(fewer trajectories and shorter evolution times). 
We use the parameters similar to those in the experiment \cite{Schauss2012},
assuming cold $^{87}$Rb atoms with the ground state 
$\ket{g} \equiv 5 S_{1/2} \ket{F=2,m_F=-2}$ and the Rydberg state
$\ket{r} \equiv n S_{1/2}$ with the principal quantum number $n$.
The resonant excitation of $\ket{r}$ is effected by a two-photon transition
via non-resonant intermediate state $\ket{e} = 5 P_{3/2} \ket{F=3,m_F=-3}$ 
with Rabi frequency $\Om = 2\pi \times 85\:$kHz (our definition 
of the Rabi frequency in $\mc{V}_{\mathrm{af}}^j$ differs from that 
in \cite{Schauss2012} by a factor of $\hlf$). 
The (population) decay rate of the Rydberg state $\ket{r}$ 
is $\Ga_r \simeq 0.075 \Om = 40\:$kHz \cite{Schauss2012}. 
The coherence relaxation $\Ga_z \simeq 0.3 \Om = 160\:$kHz on the 
transition $\ket{g} \leftrightarrow \ket{r}$ has two main contributions:
the decay from the intermediate state $\ket{e}\to \ket{g}$ with the 
rate $\sim 90\:$kHz and the two-photon laser linewidth $\sim 70\:$kHz, 
while at $T \sim 10\:$nK \cite{Schauss2012} the Doppler broadening 
is negligible ($\lesssim 10\:$kHz). 
The resulting (steady-state) excitation linewidth of $\ket{r}$ is 
$w \simeq 2\pi \times 0.5\:$MHz ($w \simeq 2\pi \times 120\:$kHz 
if $\Ga_z =0$). The corresponding blockade distance for the 
Rydberg states with $n \simeq 40-50$ ($C_6/ 2\pi \simeq 1 - 15\:
\mathrm{GHz}\:\mu\mathrm{m}^6$ \cite{rydcalc}) 
is $\db \sim 3.5-5.6\:\mu$m.

\begin{figure}[t]
\includegraphics[width=8.7cm]{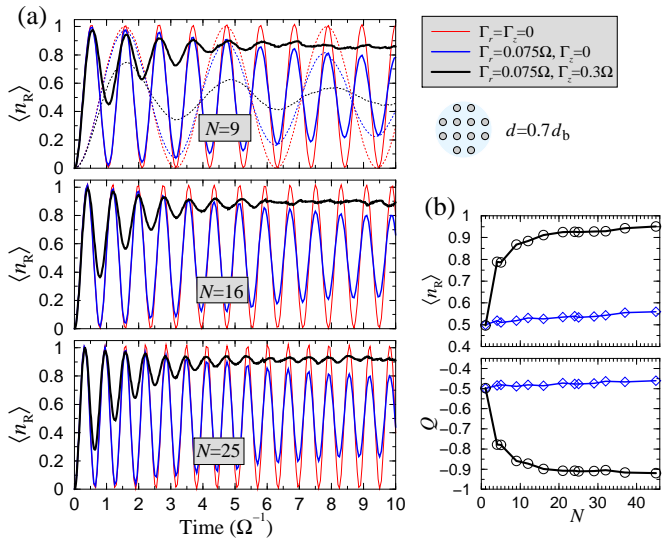}
\caption{
Dynamics and steady-state of resonantly-driven ensemble 
of $N$ atoms in a 2D lattice of diameter $d = 0.7 \db$.
(a)~Time-dependence of the number of Rydberg excitations $\expv{\nR}$
for $N=9,16,25$ and different $\Ga_{r,z}$. Thin dotted lines 
in the top graph correspond to a single atom, $N=1$.
Time is in units of $\Om^{-1} \simeq 1.87\:\mu$s.
(b)~Steady state values of $\expv{\nR}$ and $Q$ versus $N \leq 45$
for $\Ga_{z} = 0$ (blue, open diamonds) and $\Ga_{z} = 0.3 \Om$ 
(black, open circles).} 
\label{fig:SAdynss}
\end{figure}

Consider first the case of a small confinement volume, $d < \db$.
Due to the blockade effect, such a volume can accommodate at most 
one Rydberg excitation \cite{Lukin2001,Dudin2012NatPh}.
Figure~\ref{fig:SAdynss} shows the dynamics of the system and 
the steady state for different $N$. For unitary system, $\Ga_{r,z}=0$,
we observe collective Rabi oscillations of $\expv{\nR}$ between
0 and 1 with frequency $\sqrt{N} \Om$. In the presence of 
a small Rydberg state decay, $\Ga_{r} \neq 0$, these oscillations 
are slowly damped and the final steady state contains 
$\expv{\nR} \simeq \frac{1}{2}$ Rydberg excitation, 
due to the saturation of the transition 
$\ket{G} \leftrightarrow \ket{R^{(1)}}$ between the collective ground state
$\ket{G} = \ket{g_1,g_2, \ldots,g_{N}}$ and the symmetric single
Rydberg excitation state $\ket{R^{(1)}} = \frac{1}{\sqrt{N}}
\sum_j^{N} \ket{g_1,g_2, \ldots,r_j,\ldots,g_{N}}$ \cite{DPMHMF2013}
[$\expv{\nR}$ slightly larger than $\hlf$ seen in Fig.~\ref{fig:SAdynss}(b)
is due to imperfect blockade of Rydberg excitation of the outermost atoms,
$\De(0.7\db) \sim 10 w$]. Simultaneously, we have $Q \simeq -\frac{1}{2}$ 
consistent with the probabilities $\pR(0) \simeq \pR(1) \simeq 0.5$.

Coherence relaxation with the rate $\Ga_{z}$ comparable to $\Om$ leads 
to $\expv{\nR}$ approaching unity, as was also discussed in \cite{DPMHMF2013}.
The reason for this is the population of all non-symmetric single Rydberg 
excitation states of the system, which, upon saturation, leaves in the 
ground state$\ket{G}$ only a small fraction of population 
$\pR(0) \simeq 1/(N+1)$, with the result that 
$\expv{\nR} \simeq \pR(1) \simeq N/(N+1)$ \cite{Honer2011}.
From the simulations we obtain small, but not negligible, 
probabilities $\pR(n) \ll 1$ of multiple Rydberg excitations $n \geq 2$, 
again due to imperfect blockade. For the mean number of Rydberg excitations
$\expv{\nR} \gtrsim 0.9$ for $N \geq 20$, we now have $Q \simeq -0.9$, 
Fig.~\ref{fig:SAdynss}(b).

\begin{figure}[t]
\centerline{\includegraphics[width=8.7cm]{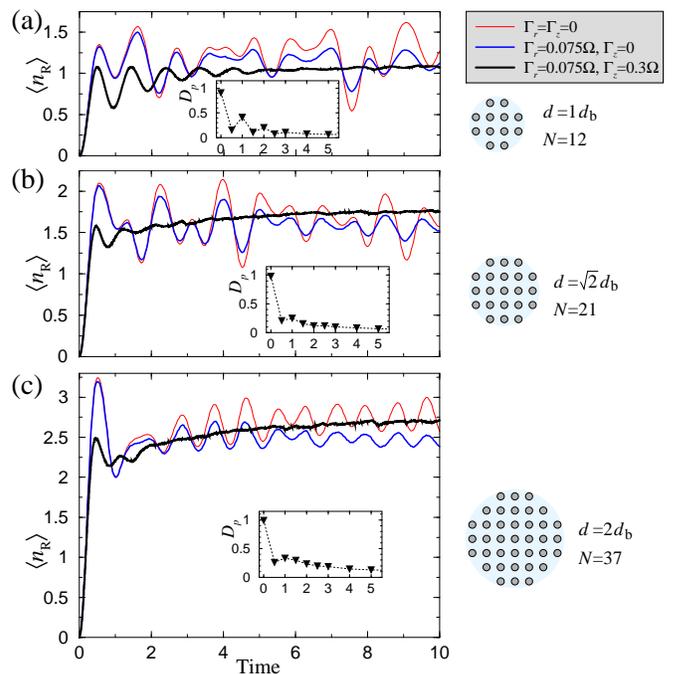}}
\caption{
Time-dependence of the number of Rydberg excitations $\expv{\nR}$
of $N=12,21,37$ atoms in the volume of size $d = (1,\sqrt{2},2)\db$,
(a,b,c), respectively. The inset in each graph shows the 
Kolmogorov distance $D_{p}$ between the probability distributions
$p_{ \{ \sigma \} }(t)$ of configurations of Rydberg excitations of 
the system at different times $t$ and $p_{ \{ \sigma \} }(\infty)$ 
in the steady-state.}
\label{fig:L12dyn}
\end{figure}

Consider next the dynamics of the system in a larger volume. 
As the system size approaches the blockade distance, $d = \db$,
the vdW interactions $\De(d) \simeq w$ between the atoms 
at the opposite sides of the confinement volume still suppress, 
but do not completely block, their Rydberg excitation, Fig.~\ref{fig:L12dyn}(a).
In effect, these interactions dephase \cite{Johnson2008,Raitzsch2008} 
the periodic Rabi oscillations of $\expv{\nR}$ even without the 
coherence relaxation, $\Ga_{z}=0$. Remarkably, $\expv{\nR}$ now 
fluctuates around 1 for both unitary and dissipative system. 
In the latter case, the coherence relaxation $\Ga_{z}= 0.3 \Om$ 
quickly damps the dynamics and the system gradually approaches 
a steady state (see below).
We observe similar behavior for larger systems, $d = \sqrt{2} \db$
and $d = 2 \db$ in Fig.~\ref{fig:L12dyn}(b) and (c), which 
of course can accommodate more Rydberg excitations, 
$\expv{\nR} \sim 1.5$ and $\expv{\nR} \sim 2.5$, respectively.

\begin{figure}[t]
\centerline{\includegraphics[width=8.5cm]{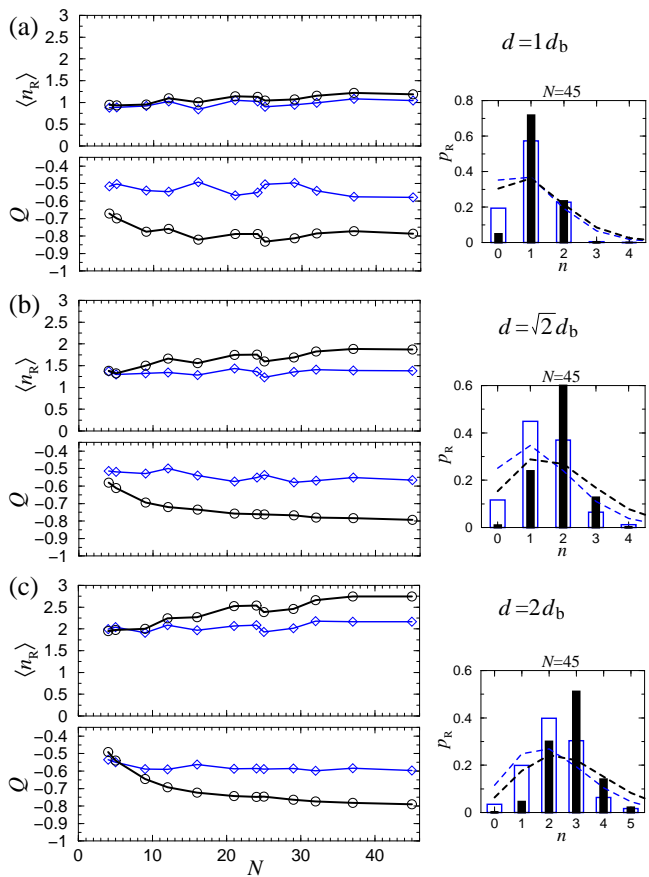}}
\caption{Steady state values of the mean number of Rydberg 
excitations $\expv{\nR}$ and the corresponding $Q$ parameter
versus the number $N \leq 45$ of atoms in the volume of size 
$d = (1,\sqrt{2},2)\db$, (a,b,c), respectively, for $\Ga_r = 0.075 \Om$,
$\Ga_z = 0$ (blue lines, open diamonds) and 
$\Ga_z = 0.3 \Om$ (black lines, open circles).  
The graphs on the right show the probability distributions 
$\pR(n)$ for $N=45$ atoms in the corresponding volume:
$\Ga_z = 0$ (blue open bars) and $\Ga_z = 0.3 \Om$ (black filled bars).
The dashed lines show the Poisson distribution for the corresponding
$\expv{\nR}$.}
\label{fig:L12ss}
\end{figure}

In the long time limit, the dissipative system equilibrates. 
Figure~\ref{fig:L12ss} shows the steady-state values of $\expv{\nR}$
and $Q$ versus the number of atoms $N$ in the confinement volume 
of diameter $d = (1,\sqrt{2},2)\db$. 
Clearly, larger volumes accommodate more Rydberg excitations, 
whose mean number $\expv{\nR}$ is, however, not directly 
proportional to $(d/\db)^2$ as one might expect for a 2D system
(and is probably true for very large volumes). This is a finite-size 
effect as the Rydberg excitations repel each other to the circular 
boundary of the confinement volume \cite{Schauss2012}. 
The non-smooth dependence of $\expv{\nR}$ (and $Q$) on $N$ is 
also due to geometric constrains of arranging an integer number 
of atoms simultaneously in the square lattice and circle which results
in different number of atoms closest to the boundaries of the volume.
Note that similarly to the small volume case, in the absence
of coherence relaxation, $\Ga_z = 0$, the mean number of
Rydberg excitations $\expv{\nR}$ in a given volume changes
little with $N$ while $Q$ stays close to $-\hlf$. In contrast,
for $\Ga_z \lesssim \Om$, $\expv{\nR}$ slowly grows with the
number of atoms $N$ (or density), while $Q \simeq -0.8$ indicates 
highly sup-Poissonian statistics of the number of Rydberg excitations, 
i.e. sharply peaked probability distribution $\pR(n)$, Fig.~\ref{fig:L12ss}.

In the experiment, one prepares all the atoms in the ground state $\ket{g}$,
applies the driving field $\Om$ for a certain time $t$ and then 
performs projective measurements $\{ \sig_{rr}^j \}$ of Rydberg 
excitations \cite{Schauss2012}. Every experimental sequence then results 
in a particular configuration $\{ \sig_{rr}^j \} \to \{0,0,1,0,\ldots \}$ 
of Rydberg excitations of the atoms. Averaging over many (typically 
several hundred) such experimental sequences yields the probabilities 
$p_{ \{ \sigma \} }(t)$ of various configurations, which are given
by the diagonal elements of the density matrix $\hrho(t)$ of the 
system in the $\{ \ket{g}, \ket{r} \}$ basis. To quantify
the dynamical approach of the system to the steady-state, 
we calculate the Kolmogorov distance 
$D_p (t) = \hlf \sum_{ \{ \sigma \} } |p_{ \{ \sigma \} }(t) - 
p_{ \{ \sigma \} }(\infty) |$, where $p_{ \{ \sigma \} }(\infty)$
are the steady-state probabilities, i.e. diagonal elements of 
$\hrho(\infty)$. The insets of Fig.~\ref{fig:L12dyn} show $D_p (t)$
at different times $t$. For a few atoms within a small volume, we observe 
initial damped oscillations and rapid approach of the system to the 
steady state configuration of Rydberg excitations, Fig.~\ref{fig:L12dyn}(a).
Increasing the number of atoms and the size of the system leads to less 
pronounced oscillations of $D_p (t)$ and slower equilibration, 
Figs.~\ref{fig:L12dyn}(b,c).
This is to be expected since, on the one hand, more atoms at various 
distances from each other result in stronger dephasing, and, on the other
hand, it takes now longer time to establish correlations between distant 
atoms which interact weakly, or negligibly, with each other. 
The similarly slow approach of a 1D lattice system to the global 
steady state was discussed in \cite{Hoening2013}.   

We note that the system attains its true steady-state on a time scale 
of $t \gtrsim 20\: \mu$s, which in an experiment can be prohibitively 
long due to the loss of atoms and the need of continuous laser irradiation.  
However, already after several $\mu$s of laser driving \cite{Schauss2012}, 
the probability distribution $p_{ \{ \sigma \} }$ of Rydberg excitation 
configurations is close to the steady-state distribution, 
with the distance between the two $D_{p} \lesssim 0.1$.


To conclude, we considered resonant Rydberg excitation of atoms 
in the presence of small population decay of the Rydberg states 
and sizable decay of atomic coherence, consistent with real
experimental situations. We analyzed the dynamics of the many-body 
system and its approach to the steady state. 
An important and perhaps counterintuitive result that emerged from
our studies is that the decay of atomic coherence, while destroying 
the quantum correlations, or entanglement, between the atoms, 
amplifies classical correlations, leading to narrower sub-Poissonian 
probability distribution of the number of Rydberg excitations quantified 
by larger negative values of the Mandel $Q$ parameter.    

We finally note that for the parameters used in Fig.~\ref{fig:L12ss}, 
the many-body steady state is not too far from the classical, for 
which the density matrix $\hrho_{\mathrm{cl}}(\infty)$ is diagonal in the 
$\{ \ket{g}, \ket{r} \}$ basis. We have calculated the trace distance 
$D_\rho = \hlf \mathrm{tr} |\hrho(\infty) -\hrho_{\mathrm{cl}}(\infty)|$
between the complete density matrix of the system $\hrho(\infty)$
and $\hrho_{\mathrm{cl}}(\infty)$, obtaining $D_\rho \simeq 0.2-0.3$. 
Increasing the relaxation rate $\Ga_z$ decreases $D_\rho$, and for 
$\Ga_z\simeq \Om$ we obtain $D_\rho \lesssim 0.1$ and nearly constant 
$Q \simeq -0.85$ versus the atom number $N \gtrsim 20$ and $d > \db$. 
The near-diagonality of the density matrix $\hrho(\infty)$, i.e. smallness 
of inter-atomic coherences, suggest that the steady state of the many-body 
system can efficiently be simulated using semiclassical methods, such as, 
e.g., rate equations or Monte Carlo sampling \cite{DPMHMF2013}. The results 
of such simulations for large 2D systems will be reported elsewhere.

\end{document}